\documentclass[twocolumn,groupedaddress,superscriptaddress,floatfix,aps]{revtex4}

\usepackage[english]{babel}
\usepackage{graphicx}
\usepackage{amsmath,bm}
\usepackage{mathrsfs}
\usepackage{amsfonts}
\usepackage{amssymb}
\usepackage[utf8]{inputenc}
\usepackage{booktabs}
\usepackage{color}
\usepackage[squaren]{SIunits}
\usepackage{multirow}

\usepackage[colorlinks=true,bookmarks=false,citecolor=blue,urlcolor=blue,pdfstartview=FitH]{hyperref}

\newcommand{\rot}[1]{\textrm{rot}(#1)}
\renewcommand{\div}[1]{\textrm{div}(#1)}

\begin{document}

\title{Stochastic band structure for waves propagating in periodic media or along waveguides}

\author{Vincent Laude}
\email{vincent.laude@femto-st.fr}
\affiliation{Institut FEMTO-ST, CNRS, Univ. Bourgogne Franche-Comt{\'e}, Besan\c{c}on, France}
\author{Maria E. Korotyaeva}
\affiliation{Institut FEMTO-ST, CNRS, Univ. Bourgogne Franche-Comt{\'e}, Besan\c{c}on, France}

\begin{abstract}
We introduce the stochastic band structure, a method giving the dispersion relation for waves propagating in periodic media or along waveguides, and subject to material loss or radiation damping.
Instead of considering an explicit or implicit functional relation between frequency $\omega$ and wavenumber $k$, as is usually done, we consider a mapping of the resolvent set in the dispersion space $(\omega, k)$.
Bands appear as as the trace of Lorentzian responses containing local information on propagation loss both in time and space domains.
For illustration purposes, the method is applied to a lossy sonic crystal, a radiating surface phononic crystal, and a radiating optical waveguide.
The stochastic band structure can be obtained for any system described by a time-harmonic wave equation.
\end{abstract}

\maketitle

\section{Introduction}

The dispersion relation is essential information to describe wave propagation~\cite{brillouinBOOK2003}.
This is especially true for structures that support wave propagation, such as waveguides and artificial crystals, including photonic~\cite{yablonovitchPRL1987,johnPRL1987,luNP2014} and phononic~\cite{kushwahaPRL1993,maldovanN2013,husseinAMR2014} crystals.
The dispersion relation gives the possible propagation modes and relates the angular frequency $\omega$ with the wavevector $\bm{k}$.
In the case of periodic media and crystals, it is termed the band structure.
It is very often obtained by looking for the eigenvalues and eigenfunctions of a matrix, in the case of finite-dimensional problems.
Finding the eigenvalues and eigenfunctions of a finite-size matrix is indeed nowadays a well mastered numerical problem~\cite{golubJCAM2000}.
Solvers are routinely used to obtain them, at least in the case of lossless waveguides and crystals.
Indeed, in the absence of loss, one often obtains a self-adjoint propagation operator, i.e. an operator satisfying Hermitian symmetry.
In this case, the Hilbert-Schmidt theorem tells us that the spectrum lies on the real line~\cite{renardyBOOK2006}.
If the operator is furthermore compact, eigenvalues are discrete and isolated (in case they are degenerate, the corresponding eigenfunctions are orthogonal).
Physically, this situation is generally implied when one plots the dispersion relation as a graph $\omega(\bm{k})$.
Both frequency and wavevector are real quantities and the dispersion relation is composed of distinct bands that can be numbered.
Note that compactness here refers to the domain of definition being finite and the operator being bounded.

\begin{figure}[!t]
\includegraphics[width=65mm]{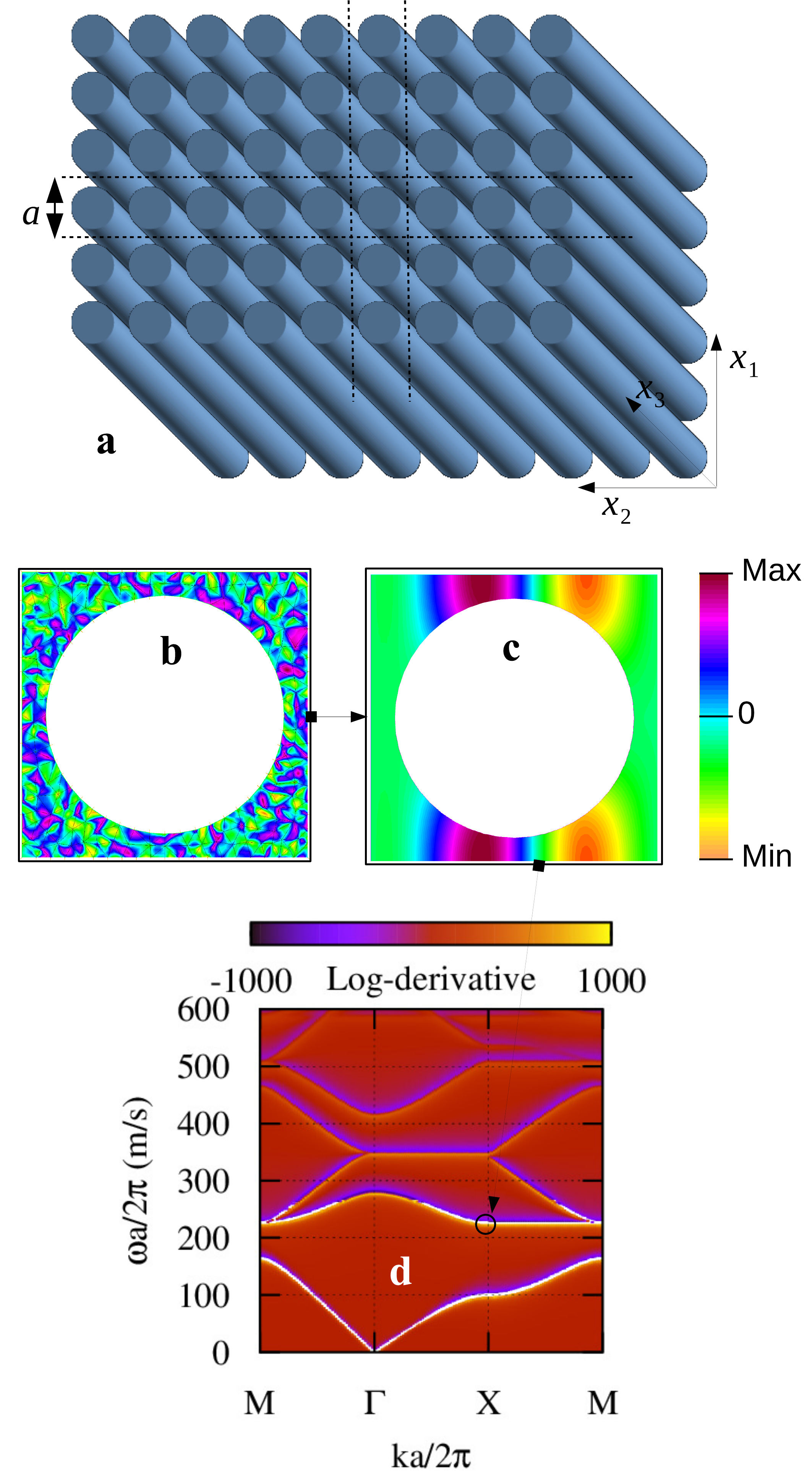}
\caption{
Application of the stochastic band structure method to a two-dimensional sonic crystal of rigid cylinders in air, with material loss taken into account.
(a) The finite-size computation domain is here a unit-cell of the crystal.
(b) A spatially random excitation is applied and the response (c) is obtained.
From the response, the stochastic band structure (d) can be plotted.
For further details see Section~\ref{sec3.1} and Fig.~\ref{fig2}.
}
\label{fig1}
\end{figure}

The situation becomes more obscure when propagation loss or some infinite dimension enters the picture.
In the case of loss, the self-adjoint property is lost.
In the case of an infinite domain of definition, the operator is no more compact.
In the latter case, there exist computational techniques to approximate the problem on a finite domain, such as the perfectly matched layer (PML)~\cite{berengerJCP1994}.
As a result, however, self-adjointness is generally lost.
Fortunately, a theorem similar to the Hilbert-Schmidt theorem still holds for compact operators in finite dimensional Hilbert spaces, but eigenvalues are complex~\cite{renardyBOOK2006}.
As a result, the bands composing the dispersion relation are not real functions anymore.
Different physical examples where this situation happens will be considered in the following of this paper: a lossy sonic crystal~\cite{moiseyenkoPRB2011}, such as the one depicted in Figure~\ref{fig1}, a phononic crystal of pillars sitting on a semi-infinite substrate~\cite{khelifPRB2010}, and an open optical waveguide.
Specifically, the goal of this paper is to obtain a generalized representation of the dispersion relation that applies to these difficult cases.
We term the method the stochastic band structure, because it relies on an analysis of the response of the propagation medium to a random excitation having a definite frequency and wavevector.
Fig.~\ref{fig1}(b) shows a typical random excitation applied to a unit-cell of a crystal.
The idea of considering the dispersion relation as a graph of discrete bands $\omega(\bm{k})$ is replaced by a response function $E(\omega, \bm{k})$ similar to a density of states.
The response itself is obtained from the solution to the forced problem, as shown in Fig.~\ref{fig1}(c).
Infinite domains are transformed to a compact domain using a PML technique implemented as a complex coordinate transform~\cite{hugoninJOSAA2005}.
As we show in Section 2, the stochastic response reveals the spectrum as the complement of the resolvent set of the operator and gives essential information on the amount of propagation loss at a particular dispersion point.
Furthermore, the response can be made almost independent of the particular realization of the random excitation.
In Section 3, we describe the application of the method to the three different situations mentioned above.
In Section 4, and before concluding the paper, we discuss the merits of the stochastic band structure method and its relations to previous methods.

\section{Theory}

Let us first recall some results from spectral theory~\cite{renardyBOOK2006}.
We place ourselves in an appropriate functional space, usually a Hilbert space.
For a bounded linear operator $A$, $R(\lambda)=(A - \lambda I)^{-1}$ is by definition the resolvant operator.
the resolvent set is the set of all complex numbers $\lambda$ such that $R(\lambda)$ exists and is bounded.
The spectrum is defined as the complement of the resolvent set in the complex plane. 
Every eigenvalue of the operator $A$ belongs to the spectrum.
Note that this definition is more general than the usual definition of eigenvalues of matrices of finite-dimensional spaces.
It also avoids the difficulties of defining the spectrum through a singular equation by considering instead its non-singular complement.

According to the Hilbert-Schmidt theorem, the spectrum of a self-adjoint operator lies on the real line and is in general a spectral combination of a point spectrum of discrete eigenvalues and a continuous spectrum. 
For compact self-adjoint operators, eigenvalues are discrete and isolated.
Eigenvalues $\lambda_n$ and eigenfunctions $\bm{e}_n$ satisfy $A \bm{e}_n = \lambda_n \bm{e}_n$.
The eigenfunction expansion theorem tells us that any function can be written $\bm{f} = \sum_n f_n \bm{e}_n$ with the eigenfunctions $\bm{e}_n$ forming an orthonormal basis.
The notation $f_n = \langle \bm{f}, \bm{e}_n \rangle$ is for the scalar product in the functional space.
As we indicated in the introduction, in case the operator is not self-adjoint, a similar theorem holds for compact operators in finite dimensional Hilbert spaces, providing the eigenvalues are considered complex.
Note that they can further be algebraically degenerated, but we will not consider this complication in this paper.
This assumption is equivalent to assuming the operator is isomorphic to a diagonalizable matrix.

Linear wave equations, including those for acoustic, elastic, and optical waves, can generally be written for time-harmonic waves as
\begin{align}
(K(\bm{k}) - \omega^2 M) \bm{u}(\omega, \bm{k}) &= \bm{f}(\omega, \bm{k})
\label{eq1}
\end{align}
where $K$ is a stiffness operator and $M$ a mass operator, $\bm{u}(\omega, \bm{k})$ is a function describing the solution in reciprocal space, and $\bm{f}(\omega, \bm{k})$ is a forcing term at a particular frequency and wavevector.
Equation \eqref{eq1} is obtained from a Fourier transform of the original wave equation over the time coordinate.
The stiffness operator $K$ is a differential operator of the space coordinates and can depend on an imposed wavevector $\bm{k}$, for waveguide and artificial crystal problems.
We can then define the resolvent operator as
\begin{align}
R(\lambda) &= (M^{-1} K - \lambda I)^{-1} 
\label{eq2}
\end{align}
with $\lambda=\omega^2$.
Note that the inversion operation of the mass operator should be understood as symbolic and is introduced for convenience; in practice there is no need to invert a matrix or operator.
The solution to Eq. \eqref{eq1} is formally
\begin{align}
\bm{u} &= R(\lambda) M^{-1} \bm{f} .
\label{eq3}
\end{align}
Introducing the eigenvalues $\lambda_n$ and eigenfunctions $\bm{e}_n$ of $M^{-1} K$, we obtain from the eigenfunction expansion theorem that
\begin{align}
u_n &= (\lambda_n - \lambda)^{-1} g_n
\label{eq4}
\end{align}
with $\bm{g} = M^{-1} \bm{f}$.
Overall, the solution is
\begin{align}
\bm{u} &= \sum_n (\lambda_n - \lambda)^{-1} g_n \bm{e}_n .
\label{eq5}
\end{align}
This equation expresses the well known fact that the solution to a linear equation is a linear combination of eigenfunctions.
The coefficients of the combination are complex Lorentzian functions or poles, centered on the eigenvalues, and are also proportional to the projection of the excitation on each eigenfunction.
When $\lambda \approx \lambda_n$, we have $\bm{u} \approx (\lambda_n - \lambda)^{-1} g_n \bm{e}_n$, i.e. the solution approaches in the limit the particular eigenfunction.
This observation leads to a practical way to obtain every eigenvalue and eigenfunction: one can explore all values of $\lambda$, i.e. the resolvent set; since the eigenvalues are isolated each pole can be isolated and identified. Of course, such a procedure would be very lengthy compared to existing eigenvalue solvers.
If we forget the idea of obtaining exactly all eigenvalues, but only wish to obtain a view of the landscape of the resolvent set with a given -- and limited -- resolution, then the method can be useful, as we illustrate in the following section.

Eq. \eqref{eq1} can be viewed as the dynamical equation obtained from the Euler-Lagrange principle with a Laplacian combining potential elastic energy, kinetic energy, and the work of the forcing term.
A Hamiltonian operator can then be defined as
\begin{align}
H &= \frac{1}{2}(K + \omega^2 M)
\end{align}
and we evaluate the response as the self-energy of the solution, or
\begin{align}
E &= \langle H \bm{u}, \bm{u} \rangle .
\end{align}
Using the eigenfunction expansion \eqref{eq5} for the solution, we have
\begin{align}
E(\lambda) &= \sum_{n,n'} \langle H \bm{e}_n, \bm{e}_{n'} \rangle g_n (\lambda - \lambda_n)^{-1} g^*_{n'} (\lambda^* - \lambda_{n'}^*)^{-1}
\end{align}
which is real and positive by construction, if $K$ and $M$ are real operators.
The response is an expansion over complex poles centered on the eigenvalues.
Close to an eigenvalue, i.e. when $\lambda \approx \lambda_n$, $E(\lambda) \approx \langle H \bm{e}_n, \bm{e}_{n} \rangle |g_n|^2 |\lambda - \lambda_n|^{-2}$.
The response has then locally a -- possibly damped -- Lorentzian shape.

In order not to miss any of the poles, it is needed that $g_n = \langle \bm{g}, \bm{e}_n \rangle \neq 0$ for any $n$.
In practice, we can consider a spatially random excitation.
If the number of degrees of freedom is large, then the probability that any $g_n = 0$ is very small.
The result can further be made almost independent of the exact value of $g_n$ by considering the log-derivative of the response
\begin{align}
\frac{\partial}{\partial \lambda} \log E = \frac{1}{E} \frac{\partial E}{\partial \lambda} .
\end{align}
Indeed, writing $\lambda - \lambda_n = \alpha + \imath \beta$, with both $\alpha$ and $\beta$ real, we have locally
\begin{align}
\frac{\partial}{\partial \alpha} \log E &\approx -2 \alpha (\alpha^2 + \beta^2)^{-1}
\label{eq10}
\end{align}
which is a real Lorentzian function.
The response is thus practically independent of the excitation.
Of course, Eq.~\eqref{eq10} is only valid close to eigenvalue $\lambda_n$ that is separated from the other eigenvalues.
In practice, it also requires that the eigenvalues are sufficiently isolated compared to the analyzing resolution.

As a whole, the stochastic response $E(\omega, \bm{k})$ contains information on the bands in the form of a continuous map in the dispersion space $(\omega, \bm{k})$.
Each band leaves a trace which is locally a Lorentzian function.
This Lorentzian function gives both the real part and the imaginary part of the local eigenvalue; in particular the width of the response informs on propagation damping in both time and space. 

\section{Results}
\label{sec3}

In this section we consider three different examples of the application of the stochastic band structure method.
In each case, a direct eigenvalue analysis would have been difficult, as we argue.

\subsection{Bloch waves in a lossy sonic crystal}
\label{sec3.1}

Let us consider a two-dimensional sonic crystal composed of steel cylinders in air, as depicted in Figure~\ref{fig1}(a).
The structure is an infinite periodic repetition of a square primitive unit-cell of length $a$, periodic along axes $x_1$ and $x_2$.
The steel cylinders are assumed infinitely long in the $x_3$ direction.
Time-harmonic waves satisfy the linear acoustic equation for pressure $p$ in air
\begin{align}
- \nabla \cdot \left( \frac{1}{\rho} \nabla p \right) - \omega^2 \frac{p}{B} &= f
\label{eq11}
\end{align}
where the mass density $\rho(\bm{r})$ and the elastic modulus $B(\bm{r})$ are functions of position and $f(\bm{r})$ is an applied forcing term.
These functions are discontinuous at the interface between matrix and inclusion.
The band structure of a sonic crystal gives the dispersion relation of its Bloch waves.
Bloch waves are in this case of the form $p(\bm{r},t)=\bar{p}(\bm{r})\exp(\imath(\omega t - \bm{k}\cdot\bm{r}))$, with $\bar{p}(\bm{r})$ the periodic part of the solution.
Because of the very large acoustic impedance mismatch between steel and air, the boundary condition on the cylinders can be approximated by $\frac{\partial p}{\partial n}=0$ (the normal acceleration vanishes on the boundary).
Appendix \ref{app1} summarizes how to transform Eq.~\eqref{eq11} to an integral equation, using the finite element method (FEM), that is then easily cast into a linear system \eqref{eq1} and to an eigenvalue system in case the applied force vanishes.

\begin{figure}[!t]
\includegraphics[width=85mm]{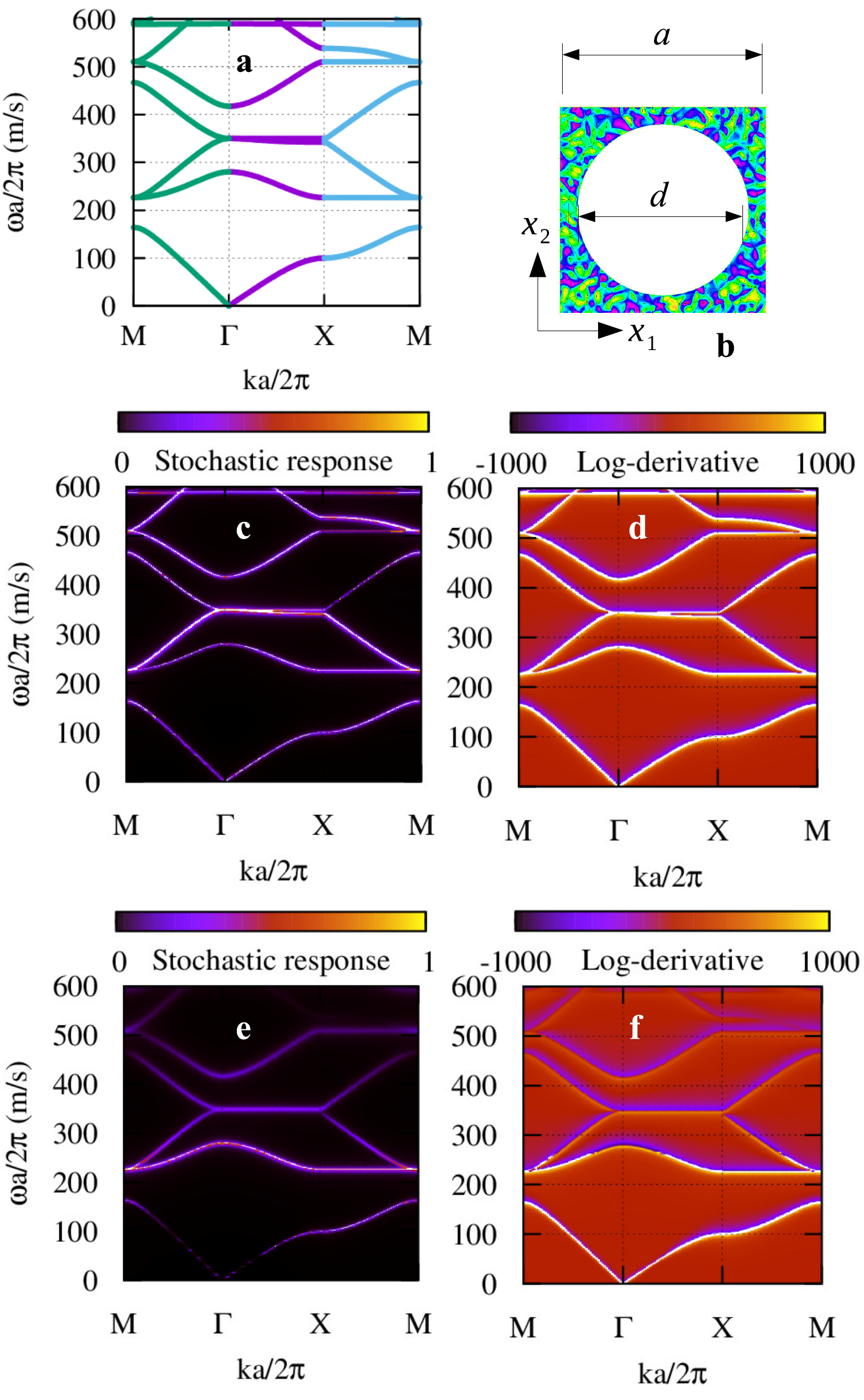}
\caption{
A two-dimensional square-lattice sonic crystal of rigid cylinders in air, with $d/a=0.8$.
(a) The band structure for the lossless crystal is obtained by solving for frequency as a function of wavevector, $\omega(\bm{k})$, along the path M-$\Gamma$-X-M of the first Brillouin zone.
(b) The stochastic excitation is applied in air as a random periodic field.
(c-d) The normalized stochastic response and its log-derivative for small loss ($\mu / B a=10^{-9}$ \meter\per\second) show mostly undamped Lorentzian functions following the bands in (a).
(e-f) The normalized stochastic response and its log-derivative for larger loss ($\mu / B a=10^{-7}$ \meter\per\second) show how damping distributes along each band.}
\label{fig2}
\end{figure}

In the absence of loss, the eigenvalue problem can be solved readily and the band structure is usually presented as reduced frequency, $\omega a/2\pi$, as a function of reduced wavenumber, $k a/2 \pi$.
The result is shown in Fig.~\ref{fig2}(a).
As the problem is self-adjoint and the unit-cell is closed (compact), eigenfrequencies are discrete and isolated.

Let us now consider material loss.
Specifically, loss is generally frequency dependent and is often modeled by a complex elastic modulus of the form $B' = B + \jmath \omega \mu$, with $\mu$ the viscosity~\cite{auldBOOK1973}.
Bloch's theorem remains valid for coefficients that depend on frequency, so we can still use the same formulation as before.
The eigenvalue problem, however, becomes nonlinear.
An alternative is to consider the complex band structure, i.e. to solve for $k$ as a function of $\omega$~\cite{laudePRB2009,romeroAPL2010}.
In this way, the spatial damping of time-harmonic waves can be obtained as a function of frequency.

In order to apply the stochastic band structure method, we apply a stochastic Bloch-Floquet excitation, $f(\bm{r}; \bm{k},\omega)=\bar{f}(\bm{r})\exp(\imath(\omega t - \bm{k}\cdot\bm{r}))$, with stochastic periodic part, to the acoustic equation.
We then explore the resolvant set as a function of $k$ and $\omega$, and obtain the stochastic band structures in Fig.~\ref{fig2}(c-f).
For illustration purposes, we have considered two arbitrary amounts of viscous damping, either $\mu / B a=10^{-9}$ \second\per\meter ~or $\mu / B a=10^{-7}$ \second\per\meter.
With these values, and for the highest reduced frequency in the band structure, the ratio $\omega \mu/B$ of the imaginary part to the real part of the elastic modulus is at most $4\times 10^{-6}$ and $4\times 10^{-4}$, respectively.
For the smaller value of viscous damping, the stochastic band structure is almost not affected and every band in the lossless band structure is visible with about the same intensity.
For the larger value of viscous damping, bands are increasingly damped with increasing frequency.

\begin{figure}[!t]
\includegraphics[width=75mm]{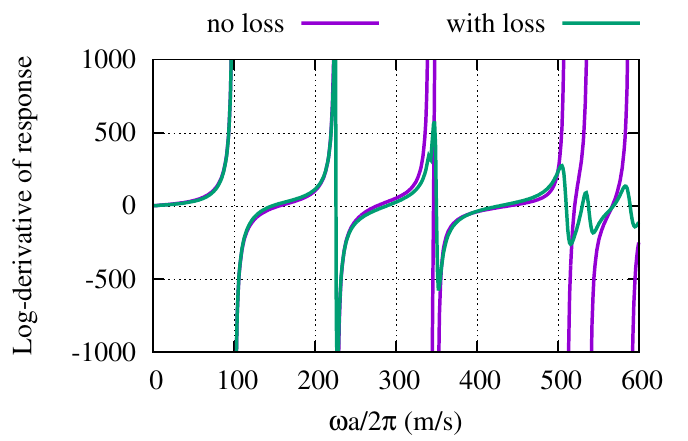}
\caption{Cross-section of the stochastic band structure of Fig.~\ref{fig2}, taken at the X point of the first Brillouin zone, without and with loss ($\mu / B a=10^{-7}$ \meter\per\second).
Temporal damping affecting the different Bloch waves is apparent.}
\label{fig3}
\end{figure}

Frequency-dependent loss can be estimated by looking at a cross-section of the stochastic band structure at constant $k$.
For instance, Fig.~\ref{fig3} shows this cross-section at the X point of the first Brillouin zone.
A superposition of damped Lorentzian functions is clearly observed.
Such vertical cross-sections reveal the amount of temporal damping for every eigenvalue.
reciprocally, considering horizontal cross-sections would reveal spatial damping of the same eigenvalues.
The stochastic band structure thus contains the information of the complex band structure, with the imaginary part of the wavenumber replaced by the width of the Lorentzian response for each band.
At the same time, it also contains information on temporal damping that is absent of the complex band structure.

\subsection{Surface Bloch waves of a phononic crystal of pillars}
\label{sec3.2}

Surface elastic wave propagation on the surface of a phononic crystal has attracted a lot of attention, partly with regards to applications to surface acoustic wave technology~\cite{olssonRST2009,benchabaneAPL2015}, to micro-electro-mechanical systems (MEMS)~\cite{mohammadiAPL2009,hsuAPL2011}, and to thermal transport~\cite{hopkinsNL2010,maldovanN2013}, but also from a fundamental point of view.
If initial questions were related to the generalization of the definition of Rayleigh surface waves to periodic media~\cite{tanakaPRB1998}, it was soon realized that surface phononic crystals must as well support the propagation of radiating guided waves~\cite{wuPRB2004,laudePRE2005}.
These waves can be described as surface excitations that are coupled with radiation modes of the substrate supporting the crystal.
Radiation modes exist in the region of the dispersion diagram called the sound cone.
By definition, the boundary of the sound cone is obtained by looking for the slowest bulk wave propagating in a given direction and along the surface of the substrate, with the direction of wave propagation being measured by the Poynting vector~\cite{laudeAPL2011}.
This procedure leads in general to an anisotropic but non dispersive velocity surface $v_{sc}(\bm{k}/|\bm{k}|)$ whose projection on the band structure looks like a cone.

The elastodynamic equation is
\begin{align}
- T_{i j,j} - \rho \omega^2 u_{i} &= f_i
\label{eq12}
\end{align}
where $\bm{u}$ is the displacement vector and $T_{i j}$ is the stress tensor.
$f_i$ are body forces and the constitutive relation of elasticity (Hooke's law) is
\begin{align}
T_{i j}  &= c_{i j k l}  S_{k l} ,
\label{eq13}
\end{align}
with $c_{i j k l}$ the elastic tensor and $S_{i j}  = \frac{1}{2} \left( u_{i,j}  +  u_{j,i} \right)$ the strain tensor.

In the general case of surface waves, and in contrast to bulk waves, there does not exist an eigenvalue equation giving the band structure.
Surface waves are instead found by looking for the zeros of a determinant of the boundary conditions, or any equivalent secular equation~\cite{wuPRB2004,laudePRE2005}.
This procedure, however, is strictly speaking limited to lossless surface waves, whose dispersion lies outside the sound cone.
Leaky guided surface waves, whose dispersion lies inside the sound cone, have been obtained by looking for minima of the boundary condition determinant~\cite{laudePRE2005}.
A difficulty is that radiation modes of the substrate have to be selected according to a partial wave selection rule.
In the general case of a finite depth phononic crystal sitting on a semi-infinite substrate, this procedure is cumbersome and the usual approach has been to consider only purely guided waves, i.e. non-radiative surface waves lying outside the sound cone~\cite{khelifPRB2010,assouarAPL2011,yudistiraAPL2012}.
An immediate drawback is that the non-radiative band structure is defined in a quite restricted sense and neglects the interaction of surface waves with bulk waves radiated in the substrate.

\begin{figure}[!t]
\includegraphics[width=85mm]{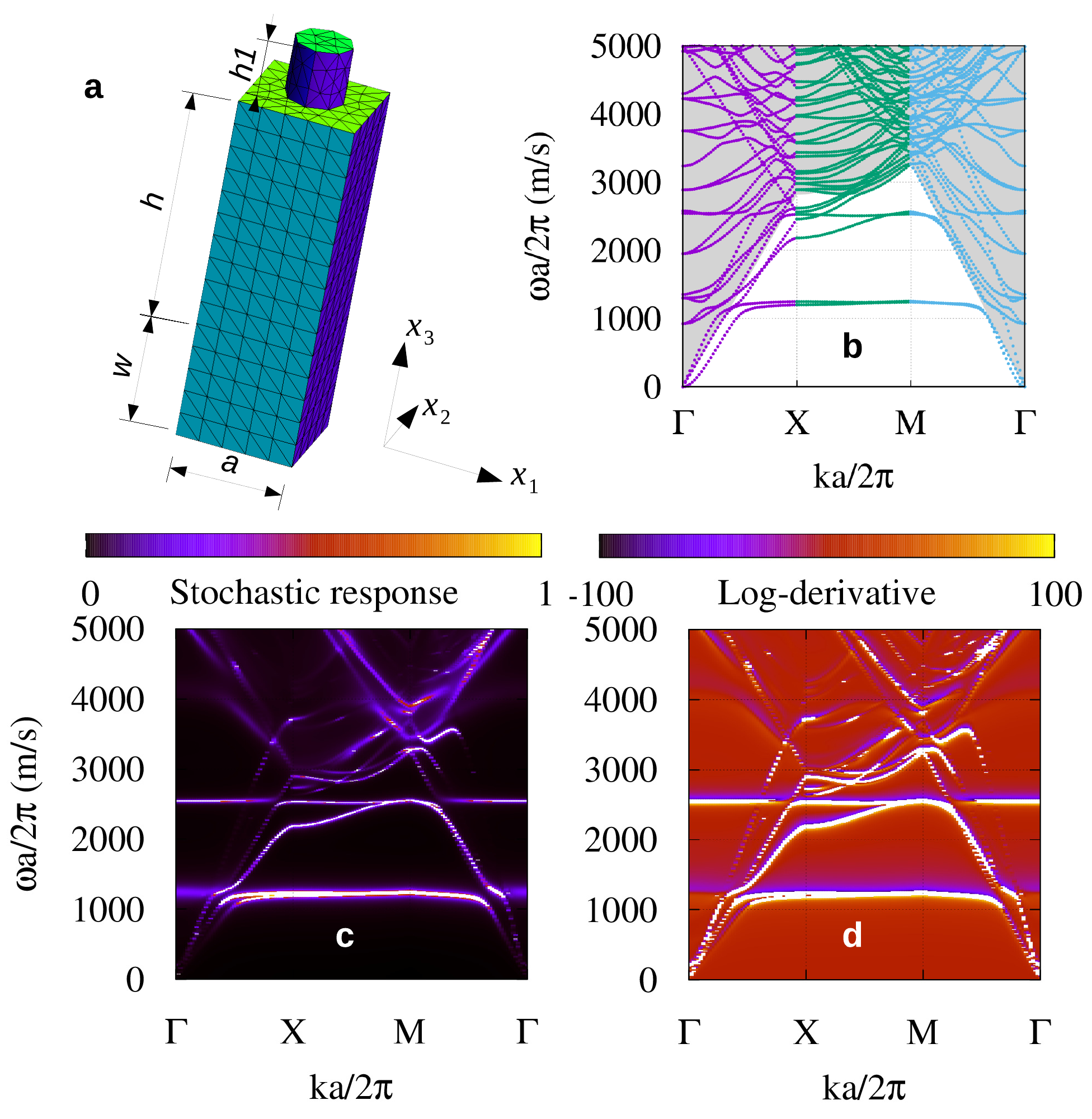}
\caption{A two-dimensional square-lattice phononic crystal of silicon pillars on a silicon substrate.
(a) The unit-cell is arbitrarily cut at a depth $h+w$ in the substrate, with $w$ the depth of a perfectly matched layer (PML) introduced to approximate radiation inside the semi-infinite substrate.
The height of the pillar is $h_1/a=1$ and the diameter is $d/a=0.5$.
(b) The classical band structure, generated by solving an eigenvalue problem with damping in the PML set off, gives the dispersion of non-radiative surface waves only outside the sound cone highlighted in gray.
(c-d) The normalized stochastic response and its log-derivative map the stochastic band structure throughout the $(\omega, k)$ dispersion plane.}
\label{fig4}
\end{figure}

As an example, let us consider the phononic crystal of pillars of Fig. \ref{fig4}.
A unit-cell of the crystal shown in Fig. \ref{fig4}(a) consists of a cylindrical pillar of the same material as the substrate, silicon.
The crystal has a square lattice with lattice constant $a$.
For obvious practical reasons, the unit-cell has to be limited to a certain depth.
Fig. \ref{fig4}(b) shows the band structure for surface guided waves computed according to the method in Ref. \cite{khelifPRB2010}.
The overlayed sound cone indicates the bands that are removed from consideration; those bands are actually strongly dependent on the substrate thickness and most of them are obviously spurious Lamb waves.

In order to apply the stochastic band structure method, the semi-infinite radiation medium has to be replaced a finite region.
A solution could be to couple the solution in a finite crystal layer with an homogeneous radiation medium.
This is however applicable only in specific cases for which the Green's function is known explicitly, such as isotropic infinite media.
Instead, we approximate numerically the semi-infinite substrate by an elastic PML, as summarized in Appendix \ref{app2}.
A Bloch-Floquet stochastic body force is applied in the layer region.

The stochastic response and its log-derivative are shown in Fig. \ref{fig4}(c-d).
Below the sound cone, the stochastic band structure is very similar to the non-radiative band structure, as expected.
Inside the sound cone, however, spurious bands are removed and radiation damping associated with each band can be easily evaluated.
In particular, the avoided crossings appearing at the intersection of local resonances of the pillars with propagating surface waves become clearly apparent.
In the classical band structure, they were damaged by interference with the spurious Lamb waves.

\subsection{Leaky guided waves in the light cone}
\label{sec3.3}

\begin{figure}[!t]
\includegraphics[width=80mm]{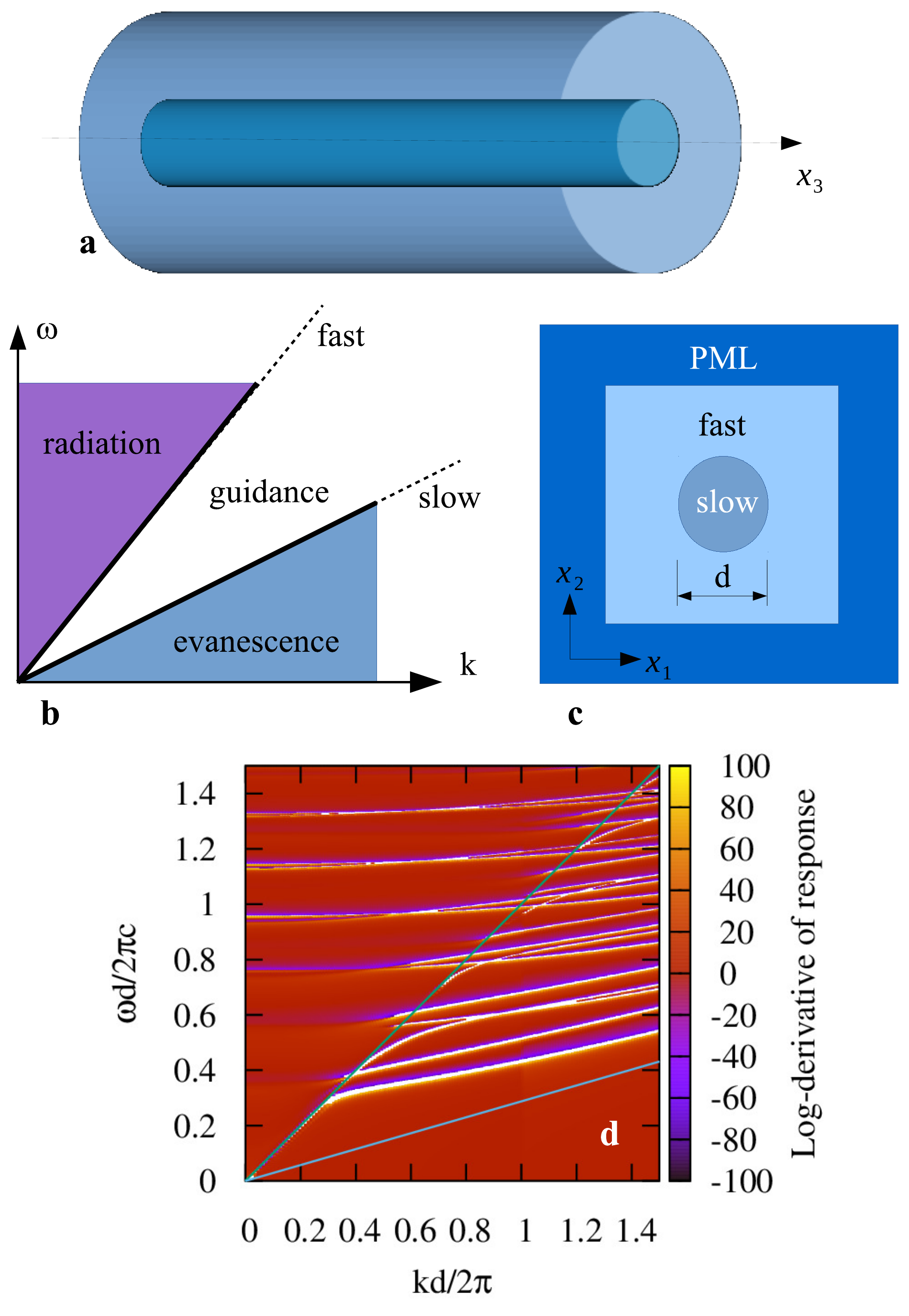}
\caption{
Optical guidance along a free-standing silicon microwire in air ($d=1$ \micro\meter).
(a) Axis $x_3$ is an invariance axis along which the wavenumber $k$ is counted.
(b) The dispersion diagram for waves guided by total internal reflection is divided in three different regions: guided waves can be either fully evanescent, guided inside the slow core, or coupled to the radiation modes of air.
(c) The computation domain for the stochastic band structure is divided between core, cladding, and a perfectly matched layer approximating radiation to infinity.
(d) The stochastic band structure maps the dispersion relation throughout the $(\omega, k)$ dispersion plane, especially informing on radiation damping inside the light cone for air.
}
\label{fig5}
\end{figure}

Waveguides are structures with one invariance axis that are able to confine propagating waves around a central core.
They have many applications in engineering, starting with the optical fiber.
There are different guidance mechanisms, including guidance provided by a photonic or a phononic band gap, but the simplest guidance mechanism is total internal reflection.
For simplicity, we will consider optical waves in the remaining of this section.
As depicted in Fig.~\ref{fig5}(a), waves can be guided in a 'slow' core surrounded by a 'fast' cladding, providing the dispersion of the guided waves lies in between the light lines of the core and the cladding~\cite{marcuseBOOK1991}.
When this condition is met, the dispersion point $(\omega, k)$ is inside the light cone of the core but outside the light cone of the cladding.
The light field is then sinusoidal in the core and evanescent in the cladding, implying confinement around the core as light propagates along the axis of the waveguide.

Now there are two more regions of dispersion space in Fig.~\ref{fig5}(a).
In the doubly-evanescent region, there are no guided solutions in optical dielectric waveguides~\cite{marcuseBOOK1991} but interface guided waves may exist in other types of systems, such as the Stoneley wave of elastic media~\cite{auldBOOK1973}.
In the radiation region, however, there exist radiation guided waves that can be regarded as a combination of waves propagating in the core with radiation modes of the infinite cladding.
Many different methods have been proposed in order to obtain radiation guided waves and their approximation with leaky waves; see for instance Ref.~\cite{huAOP2009}.
In the following we illustrate that the stochastic band structure method yields a direct mapping of dispersion and an estimate for propagation losses due to radiation.

Let us consider a free-standing silicon microwire in air, as depicted in Fig.~\ref{fig5}(b).
Maxwell's equations in dielectric media lead to the following vector wave equation for the magnetic field vector $\bm{H}$
\begin{align}
\nabla \times \left( \frac{1}{\epsilon} \nabla \times \bm{H} \right) - \frac{\omega^2}{c^2} \bm{H} &= \bm{f} ,
\label{eq14}
\end{align}
with $\epsilon$ the relative dielectric constant and $c$ the speed of light in a vacuum.
Because of invariance of the structure along axis $x_3$, solutions can be written as $\bm{\bar{H}}(x_1,x_2) \exp(\imath(\omega t - k x_3))$.
Hence the unknown becomes the modal shape $\bm{\bar{H}}(x_1,x_2)$ defined in two-dimensional transverse space.
As a result, it is enough the represent the solution on a two-dimensional mesh.
Appendix \ref{app3} gives a hybrid-mode variational formulation of guided wave optics, from which an equation system such as Eq.~\eqref{eq1} can be obtained, as well as the expression of a PML to terminate the computation domain.

\begin{figure}[!t]
\includegraphics[width=80mm]{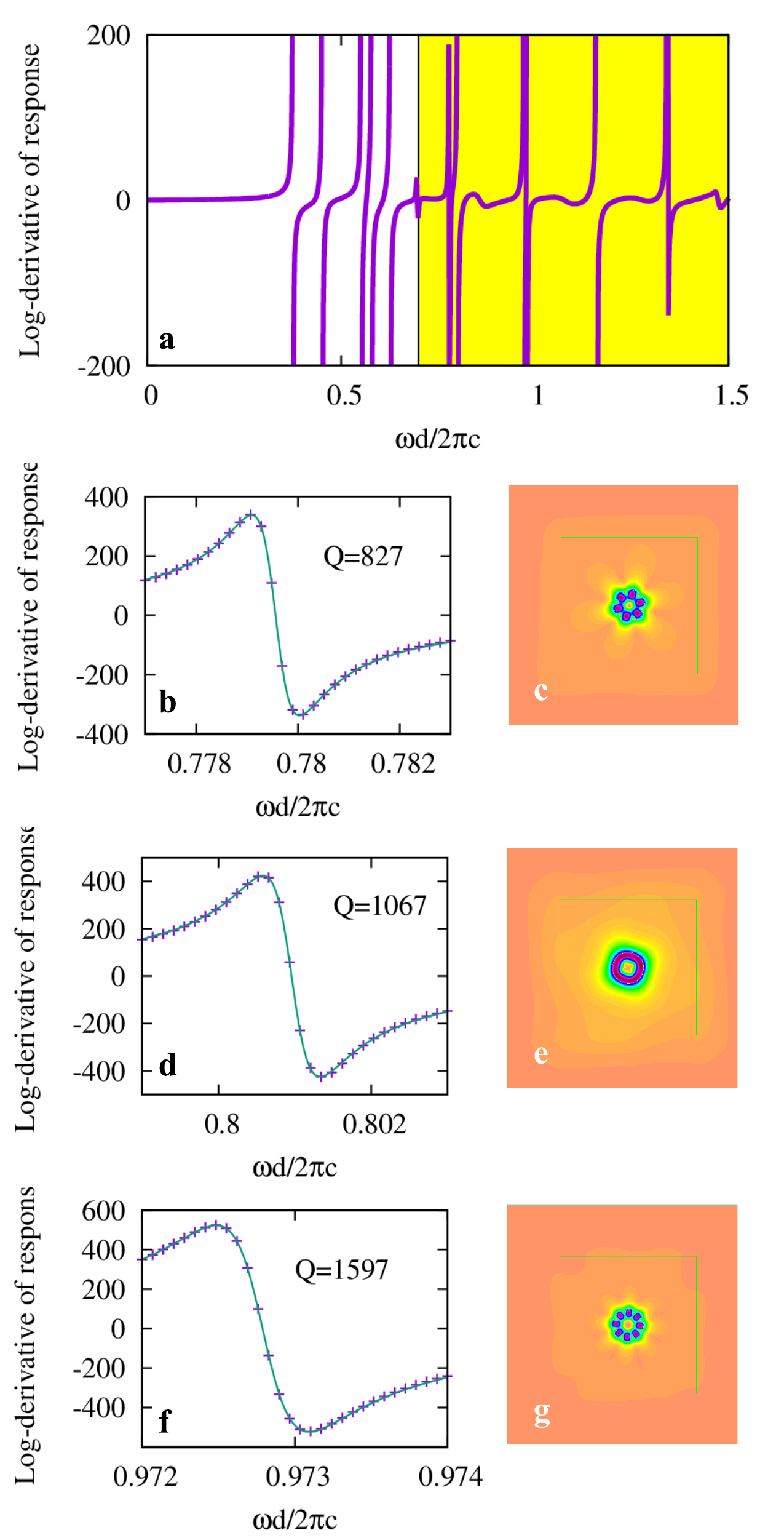}
\caption{
Optical guidance along a free-standing silicon microwire in air ($d=1$ \micro\meter).
(a) The log-derivative of the stochastic response is shown for $kd/(2\pi)=0.7$.
(b) The first damped Lorentzian resonance appearing in the light cone is fitted to the model of Eq.~\eqref{eq10}.
(c) The magnetic field distribution at the maximum of the stochastic response in (b) approximates a particular radiation guided mode.
(d-e) Same as (b-c) for the second resonance.
(f-g) Same as (b-c) for the third resonance.
}
\label{fig6}
\end{figure}

Considering guided wave stochastic excitation inside the fiber core, we obtain the stochastic band structure of Fig.~\ref{fig5}(c).
In the guidance region below the light cone for air, the usual sequence of guided modes leaves Lorentzian traces, with higher order guided modes appearing as the frequency is increased.
In the radiation region, i.e. inside the light cone for air, the stochastic band structure shows that higher order guided modes are continuously connected to transverse resonances of the microwire, with a cut-off frequency at $k=0$. 
Bands in the light cone for air are affected by radiation damping, which can be estimated from the damped Lorentzian functions.
A vertical cross-section of the stochastic band structure at $kd/(2\pi)=0.7$ is presented in Fig.~\ref{fig6}.
For frequencies under the light line the bands for guided modes are undamped.
Above the light line, however, all Lorentzian functions acquire a certain level of damping that can be attributed to coupling with radiation modes.
Each of them can be fitted individually to the model of Eq.~\eqref{eq10}, from which the quality factor can estimated as $Q \approx \beta/\Re(\lambda_n)$.
Three examples of the fitting procedure are given in Fig.~\ref{fig6}, together with the magnetic field distributions obtained at each maxima of the stochastic response.

\section{Discussion and conclusions}

As a summary, we have introduced the stochastic band structure as a generic method to obtain a mapping of the dispersion relation for wave equations.
Instead of relying on an eigenvalue problem or an equivalent root-finding method, we apply a stochastic excitation in a unit cell, with given wavenumber and frequency, and observe the response to this excitation.
Plotting the stochastic response yields a mapping of the resolvant set, of which the spectrum is the singular complement.
Close to an eigenvalue, the solution to the forced problem is proportional to the eigenfunction and the response has a simple damped Lorentzian shape, containing information on damping in both time and space.
As we have illustrated, the method (i) can be applied to frequency-dependent material loss -- it would work for $k$-dependent loss too --, (ii) takes into account radiation in an infinite or a semi-infinite medium, by combining it with a perfectly matched layer or other numerical technique, (iii) and reveals radiating guided waves and resonant modes of vibration.

Compared to eigenvalue-based methods, we don't need to assume the existence of an explicit $\omega(k)$ or $k(\omega)$ functional dispersion relation.
Instead, an implicit response $E(\omega, k)$ is obtained that is similar to a local density of states.
The full complex $(\omega, k)$ dispersion space can be explored if desired, though we have only considered its real restriction in this paper.
In the case of material loss, it would be interesting to compare further the stochastic band structure with the complex band structure~\cite{davanccoOE2007,fietzOE2011,laudePRB2009,husseinPRB2009,moiseyenkoPRB2011}.

The numerical efficiency of the stochastic band structure is rather poor, because the forced problem has to be solved for each dispersion point, and hence a great number of times in most practical problems.
It thus should not be used to replace an eigenvalue computation when the latter is possible.
The method is also best suited to small unit cells, as those used for artificial crystal and waveguide problems.

As we have shown, the result appears to be independent to a large extent from the precise random realization of the driving force, i.e. it is almost deterministic even though the generating mechanism is stochastic.
Clearly, this is possible because the stochastic excitation only appears as the right-hand-side of a linear equation; the singularities of the propagation operator filter out the solution to deliver only the eigenfunctions as maxima of the response.
By analogy with thermal fluctuations generating acoustic phonons, as observed for instance via Brillouin light scattering~\cite{carlottiAP2018} or from visualization of the vibration modes of nanomechanical resonators from thermal noise~\cite{bunchS2007,tsioutsiosNL2017}, the response has fluctuations but there is little doubt that all eigenmodes are excited for each noise realization.

We note that photonic band structures have been obtained via finite-difference time-domain (FDTD) computations~\cite{chanPRB1995}.
The FDTD method was extended to metal-dielectric~\cite{baidaPRB2006} and phononic band structures as well~\cite{tanakaPRB2000,sigalasJAP2000}.
It works by applying a spatially random excitation, with a temporal excitation, to a finite computation domain terminated by periodic boundary conditions.
By computing a Fourier transform of the solution as a function of time, a response similar to the one we have introduced is obtained.
A difference is that we work in the frequency - wavenumber domain directly, so the frequency resolution can be arbitrarily high, whereas it takes an increasingly long computation time with FDTD.
As a result, the stochastic band structure is adapted to resonant structures and very low group velocities.
Furthermore, the computation error is not growing as a function of time, as with FDTD, but is instead increasing with frequency, because the mesh captures less and less of the wave details that are of the order of the wavelength.

Finally, we suggest that the stochastic band structure can be computed for any medium supporting wave propagation as described by a time-harmonic wave equation or Helmholtz equation, which includes pressure waves in fluids, water waves, elastic waves in solids, electromagnetic waves described with Maxwell's equations -- including plasmons,-- or structures described with Schrödinger's equation.


\section*{Acknowledgments}

\noindent
This work was supported by the Agence Nationale de la Recherche through the Labex ACTION program (grant No. ANR-11-LABX-0001-01).
MK was supported by a grant from Région de Bourgogne Franche-Comté.

\appendix
\section{Mathematical models}

In this appendix, we summarize the mathematical models that were used to generate the results in Section~\ref{sec3}.

\subsection{Sonic crystal}
\label{app1}

A weak form of the linear acoustic equation \eqref{eq11} is~\cite{ihlenburgBOOOK2006}
\begin{align}
\int_\Omega \nabla q^* \cdot \left( \frac{1}{\rho} \nabla p \right)
- \omega^2 \int_\Omega q^* \frac{1}{B} p &= \int_\Omega q^* f
\end{align}
with $q$ test functions taken in the same functional space as the solution $p$.
$\Omega$ is the domain of definition, i.e. a unit cell of the sonic crystal.
Assuming a Bloch wave form for all field quantities, -- i.e. $p(\bm{r},t)=\bar{p}(\bm{r})\exp(\imath(\omega t - \bm{k}\cdot\bm{r}))$, $q(\bm{r},t)=\bar{q}(\bm{r})\exp(\imath(\omega t - \bm{k}\cdot\bm{r}))$, $f(\bm{r},t)=\bar{f}(\bm{r})\exp(\imath(\omega t - \bm{k}\cdot\bm{r}))$, -- we obtain a weak form for the periodic parts as~\cite{laudeBOOK2015}
\begin{align}
\int_\Omega (\nabla \bar{q} - \imath \bm{k} \bar{q})^* \cdot \left( \frac{1}{\rho} (\nabla \bar{p} - \imath \bm{k} \bar{p}) \right) \nonumber \\
- \omega^2 \int_\Omega \bar{q}^* \frac{1}{B} \bar{p} = \int_\Omega \bar{q}^* \bar{f}
\end{align}
with periodic boundary conditions applied on pairs of external boundaries.
When the applied force vanishes, this is an eigenvalue equation giving $\lambda=\omega^2$ as a function of $\bm{k}$.
When the applied forcing term is non zero, this is an equation system of the form of Eq.~\eqref{eq1}.
The corresponding total energy of the solution is computed as
\begin{align}
\langle H p, p \rangle &= \frac{1}{2} \int_\Omega \nabla p^* \cdot \left( \frac{1}{\rho} \nabla p \right) + \frac{1}{2} \; \omega^2 \int_\Omega p^* \frac{1}{B} p
\end{align}

\subsection{Surface phononic crystal}
\label{app2}

Since the stress and strain tensors are symmetric, we can employ the contracted notation for symmetric pairs of indices~\cite{auldBOOK1973}:
the contracted indices $I=(ij)$ and $J=(kl)$ run from 1 to 6 according to the rule $1=(11)$, $2=(22)$, $3=(33)$, $4=(23)$, $5=(13)$, and $6=(12)$.
With the definitions $T_I=T_{ij}$, and $S_J=S_{kl}$ for $I=1,2,3$ and $S_J=2S_{kl}$ for $I=4,5,6$, Hooke's law \eqref{eq13} can be written $T_I = c_{IJ} S_J$.
Considering test functions $\bm{v}$ taken in the same functional space as the solution $\bm{u}$, a weak form of the elastodynamic equation for Bloch waves is~\cite{husseinPRS2009,laudeBOOK2015}
\begin{align}
\int_\Omega S_I(\bm{v})^* c_{IJ} S_J(\bm{u}) - \omega^2 \int_\Omega \bar{\bm{v}}^* \cdot \rho \bar{\bm{u}} = \int_\Omega \bar{\bm{v}}^* \cdot \bar{\bm{f}}
\label{eqA4}
\end{align}
with
\begin{align}
S_1(\bm{u}) &= \frac{\partial \bar{u}_1}{\partial x_1} -\imath k_1 \bar{u}_1 , \\
S_2(\bm{u}) &= \frac{\partial \bar{u}_2}{\partial x_2} -\imath k_2 \bar{u}_2 , \\
S_3(\bm{u}) &= \frac{\partial \bar{u}_3}{\partial x_3} -\imath k_3 \bar{u}_3 , \\
S_4(\bm{u}) &= \frac{\partial \bar{u}_3}{\partial x_2} + \frac{\partial \bar{u}_2}{\partial x_3} -\imath (k_3 \bar{u}_2 + k_2 \bar{u}_3) , \\
S_5(\bm{u}) &= \frac{\partial \bar{u}_3}{\partial x_1} + \frac{\partial \bar{u}_1}{\partial x_3} -\imath (k_3 \bar{u}_1 + k_1 \bar{u}_3) , \\
S_6(\bm{u}) &= \frac{\partial \bar{u}_2}{\partial x_1}  + \frac{\partial \bar{u}_1}{\partial x_2} -\imath (k_2 \bar{u}_1 + k_1 \bar{u}_2) .
\end{align}
Eq.~\eqref{eqA4} is used to solve an eigensystem in case the forcing term vanishes or to obtain an equation system of the form of Eq.~\eqref{eq1} for the stochastic band structure.
The total energy of the solution is computed as
\begin{align}
\langle H \bm{u}, \bm{u} \rangle &= \frac{1}{2} \int_\Omega S_I(\bm{u})^* c_{IJ} S_J(\bm{u}) \nonumber \\
&+ \frac{1}{2} \omega^2 \int_\Omega \bm{u}^* \cdot \rho \bm{u}
\end{align}

The perfectly matched layer (PML) is next introduced to transform the infinite problem into a finite problem.
The idea is to seek a solution to the dynamical equations by using a coordinate transform from a complex infinite space, that admits evanescent waves as eigenfunctions instead of plane waves, to the real finite space~\cite{hugoninJOSAA2005,zschiedrichJCAM2006}.
Given coordinates $\bm{x}$ of real space, we introduce coordinates $\bm{y}$ of complex space via a transform $y_i=y_i(\bm{x})$.
Upon introducing the Jacobian matrix
\begin{equation}
J_{ij}=\frac{\partial y_i}{\partial x_j}
\end{equation}
we can rewrite the integrals of the variational formulation.
In an integral, the integration element changes proportionally to $\det(J)$.
Consider a function $u(\bm{x})=\tilde{u}(\bm{y})$.
The gradient of the displacement vector transforms as
\begin{equation}
\nabla \tilde{u} = \frac{\partial \tilde{u}}{\partial y_i} = \frac{\partial x_j}{\partial y_i} \frac{\partial u}{\partial x_j} = J^{-t} \nabla u.
\end{equation}
The inverse Jacobian has elements $J^{-1}_{ij} = \frac{\partial x_i}{\partial y_j}$ and $(...)^t$ denotes the transposition operator.

In the case of the elastodynamic equation, the weak form becomes~\cite{laudeBOOK2015}
\begin{align}
\int_\Omega S_I(\bm{v})^* c_{IJ} S_J(\bm{u}) \det(J) &- \omega^2 \int_\Omega \bar{\bm{v}}^* \cdot \rho \bar{\bm{u}} \det(J) \nonumber \\
&= \int_\Omega \bar{\bm{v}}^* \cdot \bar{\bm{f}}
\end{align}
with the modified definition of the strains
\begin{align}
S_{1}(\bm{u}) &= J^{-1}_{m,1} u_{1,m}, \\
S_{2}(\bm{u}) &= J^{-1}_{m,2} u_{2,m}, \\
S_{3}(\bm{u}) &= J^{-1}_{m,3} u_{3,m}, \\
S_{4}(\bm{u}) &= J^{-1}_{m,2} u_{3,m} + J^{-1}_{m,3} u_{2,m}, \\
S_{5}(\bm{u}) &= J^{-1}_{m,1} u_{3,m} + J^{-1}_{m,3} u_{1,m}, \\
S_{6}(\bm{u}) &= J^{-1}_{m,1} u_{2,m} + J^{-1}_{m,2} u_{1,m},
\end{align}
where summation over $m$ is implied and $u_{i,m}=\frac{\partial \bar{u}_i}{\partial x_m} -\imath k_m \bar{u}_i$.

In practice we use the following coordinate transform for the SAW problem, for $x_3<-h$,
\begin{align}
y_3 &= x_3 + \frac{i}{\omega} \int_{-h}^{x_3} \sigma(s) \mathrm{d}s
\end{align}
with $\sigma(s) = \beta |s + h| / w^2$, where $w$ is the PML width and $\beta$ is a numerical coefficient whose value is tuned to optimize absorption.
In Fig.~\ref{fig4}, the values $h=2a$ and $w=a$ were used.

\subsection{Dielectric optical waveguide}
\label{app3}

A weak form of the guided-wave optical equation~\eqref{eq14} is~\cite{jinBOOK2002}
\begin{align}
\int_\Omega \frac{1}{\epsilon} \left( \rot{\bar{\bm{H}}'} \rot{\bar{\bm{H}}} + \div{\bar{\bm{H}}'} \div{\bar{\bm{H}}} + k^2 \bar{\bm{H}}' \cdot \bar{\bm{H}} \right) \nonumber \\
+ \int_{\delta\Omega} \bar{H}'_n \div{\bar{\bm{H}}} \left[ \frac{1}{\epsilon} \right]
- \frac{\omega^2}{c^2} \int_\Omega \bar{\bm{H}}' \cdot \bar{\bm{H}} = \int_\Omega \bar{\bm{H}}' \cdot \bar{\bm{f}} .
\label{eqA22}
\end{align}
This expression is obtained by keeping as unknowns only the first two components of $\bm{H}$, i.e. $\bm{H} = (H_1, H_2)$, since the third component is set by the auxiliary Maxwell equation $\nabla \cdot \bm{H} = 0$.
Here we use the transverse divergence $\div{\bar{\bm{H}}} = \bar{H}_{1,1} + \bar{H}_{2,2}$ and transverse rotational $\rot{\bar{\bm{H}}} = \bar{H}_{2,1} - \bar{H}_{1,2}$, $\bar{H}'_n$ is the normal component of $\bar{\bm{H}}'$ at the boundary $\delta\Omega$, and $\left[ \frac{1}{\epsilon} \right]$ denotes the jump of the permittivity.
Note that the boundary integral appears because on the non continuity of the electric field at the interface between different dielectric media.
In Fig.~\ref{fig5}, the boundary $\delta\Omega$ is the interface between silicon and air.

The total energy of the solution is computed as
\begin{align}
E = & \frac{1}{2} \int_\Omega \frac{1}{\epsilon} \left( \rot{\bar{\bm{H}}} \rot{\bar{\bm{H}}} \right. \nonumber \\
& \left. + \div{\bar{\bm{H}}} \div{\bar{\bm{H}}} + k^2 \bar{\bm{H}} \cdot \bar{\bm{H}} \right) \nonumber \\
& + \frac{1}{2} \int_{\delta\Omega} \bar{H}_n \div{\bar{\bm{H}}} \left[ \frac{1}{\epsilon} \right] \nonumber \\
& + \frac{\omega^2}{2c^2} \int_\Omega \bar{\bm{H}} \cdot \bar{\bm{H}} .
\label{eqA23}
\end{align}

Again, Eq.~\eqref{eqA22} leads to an equation system of the form of Eq.~\eqref{eq1}.
A PML is constructed as in the elastic case from a complex coordinate transformation, now along both axes $x_1$ and $x_2$.
The integrands of the first and the third integrals in Eq.~\eqref{eqA22} are multiplied with $\det(J)$, while the definition of the transverse divergence and rotational become
\begin{align}
\div{\bar{\bm{H}}} &= J^{-1}_{m,1} \bar{H}_{1,m} + J^{-1}_{m,2} \bar{H}_{2,m}, \\
\rot{\bar{\bm{H}}} &= J^{-1}_{m,1} \bar{H}_{2,m} - J^{-1}_{m,2} \bar{H}_{1,m},
\end{align}
where summation over $m$ is implied.
\vfill


\end{document}